\documentclass[aps,prl,preprint,groupedaddress]{revtex4-1}
\usepackage{graphics}%

\begin{document}


\title{Identification of main contributions to conductivity of epitaxial InN}


\author{T.A. Komissarova}
\email[komissarova@beam.ioffe.ru]{ }
\author{V.N. Jmerik}
\author{S.V. Ivanov}
\affiliation{Ioffe Physical Technical Institute of the Russian Academy of Sciences, 26, Polytekhnicheskaya Str., 
St Petersburg 194021, Russia}
\author{O. Drachenko}
\affiliation{Institute of Ion-Beam Physics and Materials Research and Dresden High Magnetic Field Laboratory (HLD), Helmholtz-Zentrum Dresden-Rossendorf (HZDR), 01314 Dresden, Germany}
\author{X. Wang}
\affiliation{State Key Laboratory of Artificial Microstructure and Mesoscopic Physics, School of Physics, Peking University, Beijing 100871, People's Republic of China}
\author{A. Yoshikawa}
\affiliation{Graduate School of Electrical and Electronic Engineering, Chiba University, 1-33 Yayoi-cho, Inage-ku, Chiba 263-8522, Japan}


\date{\today}

\begin{abstract}
Complex effect of different contributions (spontaneously formed In nanoparticles, near-interface, surface and bulk layers) on electrophysical properties of InN epitaxial films is studied. Transport parameters of the surface layer are determined from the Shubnikov-de Haas oscillations measured in undoped and Mg-doped InN films at magnetic fields up to 63 T. It is shown that the In nanoparticles, near-interface and bulk layers play the dominant role in the electrical conductivity of InN, while influence of the surface layer is pronounced only in the compensated low-mobility InN:Mg films.
\end{abstract}


\maketitle

\section{I. Introduction}

Presently electronic properties of InN have been a subject of intense studies. Despite the predicted extraordinary electron transport parameters among other III-Nitrides, the experimentally determined Hall concentration $n_H$ in the InN films has been still rather high (10$^{18}$$\div$10$^{19}$ cm$^{-3}$), whereas the Hall mobility $\mu_H$ (100$\div$2370 cm$^2$/Vs) is significantly lower than expected \cite{1}. Recently the existence of a surface accumulation layer in InN films has been confirmed by the angle-resolved photoemission spectroscopy \cite{2}, the high-resolution electron-energy-loss spectroscopy (HREELS) \cite{3}, and the electrolyte CV (ECV) measurements \cite{4}. It has been found that the surface state density of the accumulation layer lays in the range (2$\div$6)$\times$10$^{13}$ cm$^{-2}$, and its thickness ranges within 5$\div$10 nm. It is currently accepted that the surface accumulation layer prevents proper measurements of the transport parameters of n-type and especially p-type bulk InN layers (see \cite{1} for references), although no strict experimental evidences for that have been provided. Under this assumption, the  $n$ and $\mu$ values of the bulk of InN films were calculated in numerous papers by using the model of two parallel layers \cite{5}. However the HREELS and ECV measurements provide only the values of electron concentration and thickness for the surface layer, while knowledge of the surface electron mobility is necessary for accurate evaluation of the accumulation layer effect on the electrical measurements. Different values of the electron mobility have been assigned to the surface layer in InN films, e.g. (100-200) cm$^2$/Vs \cite{6} and (700-800) cm$^2$/Vs \cite{7}. Some researchers suggested almost equal mobilities of bulk and surface electrons \cite{8}. Besides, the values of surface electron density and thickness of the accumulation layer taken from the HREELS \cite{3} and ECV \cite{4} data were applied usually in assumption of universality of the accumulation layer parameters for all InN films, which was not confirmed experimentally. Therefore, the lack of understanding of electrical processes in InN films and influence of different conducting channels on the electrical measurements, as well as the unknown transport parameters of different layers comprising InN films lead to substantial difficulties in determination of electrical properties of the bulk InN layers and their control during epitaxial growth.

This paper reports on comprehensive studies of the electrical properties of undoped and Mg-doped InN films grown by plasma-assisted molecular beam epitaxy (PA MBE). Four different contributions to the conductivity of InN: spontaneously formed In nanoparticles, surface and near-interface layers, and bulk semiconductor matrix of InN, have been detected and identified using electrical measurements in high magnetic fields. The transport parameters of surface and bulk InN layers have been determined independently. The difference between the Hall electron concentration and that determined from the Shubnikov-de Haas oscillations has been explained by the influence of the near-interface layer.

\section{II. Samples and experimental technique}

The N-polar InN films were grown by PA MBE under different stoichiometric conditions on GaN buffer layers deposited by MBE on c-sapphire substrates. In addition to the undoped InN films, Mg-doped ones grown in Chiba University \cite{5} have been studied. The magnesium concentration in the 0.6-$\mu$m-thick InN:Mg layers was varied within [Mg] = 1.3$\times 10^{17} - 6\times 10^{18}$  cm$^{-3}$, with the highest value being in the range where the surface accumulation layer is expected to hide the bulk p-type conductivity layer \cite{5}. The magnetic field dependences of the resistivity ($\rho$) and Hall coefficient ($R_H$) were measured in the temperature range of 4.2 -50 K in pulsed magnetic fields up to 63 T in two configurations $\bf{B}$$\parallel$$\bf{c}$-axis and $\bf{B}$$\perp$$\bf{c}$-axis.

\section{III. Experimental results and discussion}

The first important contribution to the conductivity of the InN epitaxial films relates to the influence of the In nanoparticles spontaneously formed during PA MBE growth due to the extremely low In-N binding energy \cite{9,10}. Recently it has been shown that presence of the In inclusions in InN films results in the abnormal magnetic-field dependence of the Hall coefficient RH and strong magnetoresistance effect \cite{11,12}. In this case, the electron concentration and mobility of the InN semiconductor matrix can be determined only from fitting the magnetic-field dependence of $R_H$ in the frames of the model taking into account presence of the highly conductive In nanoparticles  \cite{11}. $R_H$ has been found to increase with $B$ for all the investigated InN films. Therefore, the values of electron concentration $n_m$ and mobility $\mu_m$ of the InN semiconductor matrix were calculated from the $R_H$ vs $B$ dependence (Table I, columns 4,5). Undoped sample C443 differs from E974 by the higher defect density causing the larger carrier concentration and lower mobility in the InN matrix.

 \begin{table}[h!]|
\caption{ \label{t1} Experimentally determined parameters of the investigated InN films.}
 \begin{ruledtabular}
 \begin{tabular}{cccccccccc}
$\bf 1$&$\bf 2$&$\bf 3$&$\bf4$&$\bf 5$&$\bf 6$&$\bf 7$&$\bf 8$&$\bf 9$&$\bf 10$\\
\hline
Sample& d,& [Mg],& $n_m$,& $\mu_m$,&$\tau_t^{(1)}$,&$n_{SdH}^{(1)}$,&$\tau_q^{(1)}$,&$n_{SdH}^{(2)}$,&$\tau_q^{(2)}$,\\
No.&nm&cm$^{-3}$&cm$^{-3}$&cm$^2$/Vs&c&cm$^{-3}$&c&cm$^{-2}$&c\\
\hline
C443&1000&0&8.7$\times10^{18}$&1400&5.0$\times10^{-14}$&6.2$\times10^{18}$&2.0$\times10^{-14}$&
1.8$\times10^{13}$&9$\times10^{-15}$\\
E974&540&0&2.8$\times10^{18}$&2000&5.6$\times10^{-14}$&1.5$\times10^{18}$&2.9$\times10^{-14}$&
2.1$\times10^{13}$&1$\times10^{-14}$\\
E978&630&1.3$\times10^{17}$&2.1$\times10^{18}$&2000&7.3$\times10^{-14}$&1.6$\times10^{18}$&3.3$\times10^{-14}$&
1.1$\times10^{13}$&1$\times10^{-14}$\\
E980&650&1.1$\times10^{18}$&2.6$\times10^{18}$&900&3.1$\times10^{-14}$&1.8$\times10^{17}$&3.0$\times10^{-14}$&
2.5$\times10^{13}$&1$\times10^{-14}$\\
E981&630&6.0$\times10^{18}$&8.4$\times10^{17}$&600&2.1$\times10^{-14}$&3.3$\times10^{17}$&2.3$\times10^{-14}$\\
\end{tabular}
 \end{ruledtabular}
 \end{table}

\begin{figure}
\includegraphics{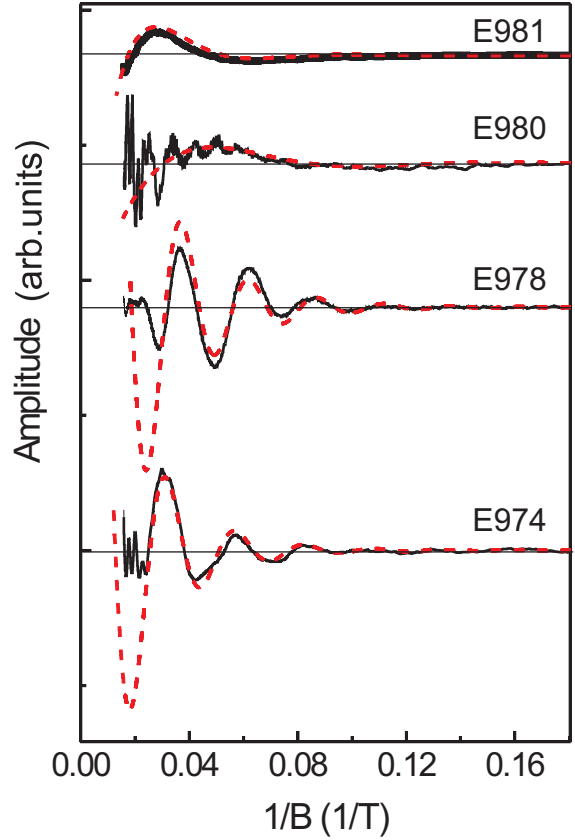}%
\caption{ \label{f1} (Color online) Experimental dependences of the oscillatory part of resistivity for different InN films (black curves) and their approximations obtained using equation (1) (dashed gray curves (online: dashed red curves)). $\bf{B}\parallel\bf{c}$-axis configuration was used. }
\end{figure}

Observation of the Shubnikov-de Haas (SdH) oscillations in the InN films allowed us to separate the contributions of the surface, near-interface and bulk layers to the conductivity of the InN matrix. The only single sets of SdH oscillations were observed in magnetic fields up to 30T (Fig.1). The period of the oscillations was the same for the $\bf{B}\parallel\bf{c}$-axis and $\bf{B}\perp\bf{c}$-axis configurations, which enables one to attribute these SdH oscillations to bulk InN layer with the typical thickness and the lateral grain size larger than the electron cyclotron orbit $\lambda$  (10-30 nm). In case of the quadratic dispersion law, the oscillatory component of the resistivity is expressed as follows \cite{13}
\begin{equation}\label{eq:1}
\Delta\rho(B)\propto\sqrt{\frac{\hbar\omega_c}{2E_F}}cos\left(\frac{\pi}{2}\frac{gm^*}{m_0} \right)\frac{2\pi^2k_BT/\hbar\omega_c}{sinh(2\pi^2k_BT/\hbar\omega_c)}exp\left(-\frac{2\pi^2k_BT_D}{\hbar\omega_c}\right)cos\left(\frac{2\pi E_F}{\hbar\omega_c}-\varphi \right)
\end{equation}
where $\omega_c=eB/m^*$ is the cyclotron frequency, $E_F$ is the Fermi energy,  $g$ is the Lande g-factor, $T_D={\hbar}/{2\pi k_B\tau_q}$ is the Dingle temperature, $\tau_q$ is the single-particle relaxation time (the quantum relaxation time), the phase $\varphi$ is a variable parameter. Knowledge of the cyclotron electron effective mass $m^*$ is necessary for approximation of the experimental curves. The $m^*$ values of about 0.05$m_0$ and 0.065$m_0$ were obtained from the temperature dependences of the SdH oscillation amplitude for undoped (E974) and slightly doped (E978) InN films, respectively. These values correspond well to the recently published data ($m^*$=0.062$m_0$) \cite{14}, which allows us to use $m^*$ = 0.062$m_0$ for the rest Mg-doped samples and 0.064$m_0$ for c443 sample. The concentrations of quantized electrons $n_{SdH}^{(1)}$ and $\tau_q^{(1)}$  (Table I, columns 7,8) have been defined from the approximation of the experimental dependences in Fig. 1, using Eq. (1).

At $B$ above 30T, the second sets of the SdH oscillations with smaller periods were observed (Fig. 1). These oscillations disappear in the $\bf{B}\perp\bf{c}$-axis configuration (Fig. 2), which indicates the two-dimensional (2D) nature of the conductivity channel having the thickness $d_s$ less than $\lambda\sim$20 nm. It is reasonable to assume that the surface accumulation layer serves as this 2D layer. The values of the two-dimensional carrier density of the quantized electrons $n_{SdH}^{(2)}$  (Table I, column 9) have been calculated from the oscillations periods and turned out to be different for different InN films, but all fell into the range of the published surface carrier densities  \cite{{2},{3},{4}}.

\begin{figure}
\includegraphics{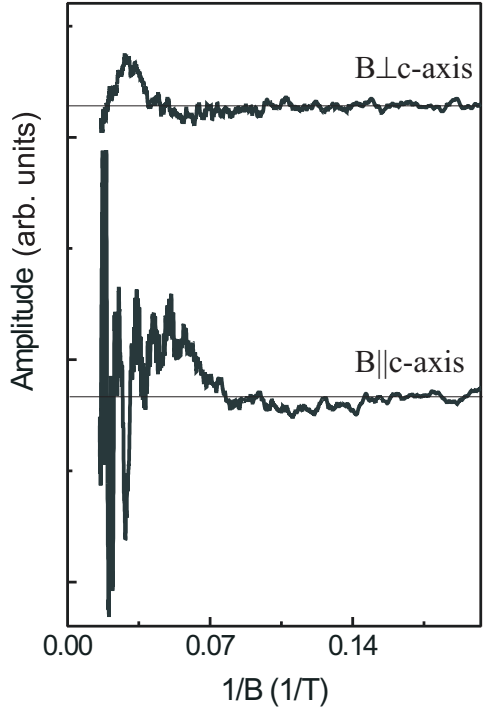}%
\caption{ \label{f2} Experimental dependences of the oscillatory part of resistivity for E980 sample in two configurations: $\bf{B}\parallel\bf{c}$-axis and $\bf{B}\perp\bf{c}$-axis.}
\end{figure}

To estimate the influence of the surface accumulation layer on Hall measurements of transport parameters of the bulk InN it is necessary to know the values of electron mobility in the surface and bulk layers. The electron mobility depends on the transport relaxation time $\tau_t$. Since the SdH oscillations corresponding to the surface layer appear at 30T, one can estimate the minimum quantum relaxation time of the surface electrons $\tau_q^{(2)}$ as $\sim1\times10^{-14}$c, having in mind the conditions of observation of SdH oscillations $\hbar/\tau_q<\hbar\omega_c$. 
In case of the ionized defect scattering, which is obviously essential for the surface 
accummulaion layer, the $\tau_t$ and $\tau_q$ values differ from each other and are related by the equation \cite{15}
\begin{equation}\label{eq:4}
\tau_t\cong\left(\frac{E_F}{\hbar} \right)^{1/2}\tau_q^{3/2}.
\end{equation}
Using the values of $\tau_q^{(2)}$ and $m^*$ = 0.09$m_0$ for the surface electrons, one can estimate  $\tau_t^{(2)}$  from Eq. (2) and calculate the mobility $\mu_{calc}^{(2)}$  which has been found to be in the range of (400-600 cm$^2$/Vs) for different InN films (Table II, column 5). The $m^*$ value has been estimated from the high surface electron density taking into account the conduction band non-parabolicity \cite{16}.

It is worth noting that the experimental values of the quantum relaxation times of bulk electrons $\tau_q^{(1)}$  and their transport relaxation times $\tau_t^{(1)}$  (Table I, columns 6,8), the latter being estimated from $\mu_m$ (Table I, column 5), differ strongly for the undoped (c443 and E974) and slightly Mg-doped (E978) InN films. This indicates that the ionized defect scattering is one of the dominant scattering mechanisms in the bulk layers as well. It enables one to define the real mobility of bulk electrons  $\mu_{calc}^{(1)}$ for these samples (Table II, column 3), using Eq. (2).

Then it becomes possible to estimate the influence of the surface accumulation layer on the measurements of transport parameters of the bulk InN layer using the model of two parallel layers. The calculated values of electron concentration $n_p$ and mobility $\mu_p$ in the case of two parallel layers are presented in Table II (columns 8,9), the thickness of the surface layer being taken as 10 nm. One can observe the negligible effect of the surface accumulation layer on Hall measurements of high-mobility undoped and slightly Mg-doped InN films. The effect of the surface layer on electrical measurements becomes pronounced only in the case of InN films with low concentration of the bulk electrons and low mobility, e.g. in the significantly compensated Mg-doped InN film (E980).

 \begin{table}[h!]|
\caption{ \label{t1} Calculated parameters of the constituent layers in the investigated InN films.}
 \begin{ruledtabular}
 \begin{tabular}{ccccccc}
$\bf 1$&$\bf 2$&$\bf 3$&$\bf4$&$\bf 5$&$\bf 6$&$\bf 7$\\
\hline
Sample&$n_{SdH}^{(1)}$,&$\mu_{calc}^{(1)}$,&$n_{SdH}^{(2)}$,&$\mu_{calc}^{(2)}$,&$n_p$,&$\mu_p$\\
No.&cm$^{-3}$&cm$^2$/Vs&cm$^{-2}$&cm$^2$/Vs&cm$^{-3}$&cm$^2$/Vs\\
\hline
C443&6.2$\times10^{18}$&1320&1.8$\times10^{13}$&450&6.2$\times10^{18}$&1310\\
E974&1.5$\times10^{18}$&2070&2.1$\times10^{13}$&560&1.6$\times10^{18}$&1970\\
E978&1.6$\times10^{18}$&1730&1.1$\times10^{13}$&400&1.6$\times10^{18}$&1700\\
E980&1.8$\times10^{17}$&900&2.5$\times10^{13}$&610&5.4$\times10^{17}$&730\\
\end{tabular}
 \end{ruledtabular}
 \end{table}

One should mention that no one of the Mg-doped films under study showed hole conductivity in the bulk layer. However, the observed high compensation of the bulk of the investigated Mg-doped InN films with[Mg] $>1\times10^{18}$cm$^{-3}$ allows us to believe that Mg doping of thicker InN films with lower residual electron concentration in the matrix could result in successful achieving of p-type conductivity.

Finally, the observed difference between the concentration of bulk quantized electrons $n_{SdH}^{(1)}$ and Hall concentration in the InN matrix $n_m$ (Table I, columns 4,7) cannot be explained only by taking into account existence of the surface layer (compare $n_m$ and $n_p$). It indicates that there exist non-quantized electrons in the InN matrix. It is well known that due to large lattice mismatch between an InN film and a GaN buffer layer the near-interface InN layer contains as usual much higher density of threading dislocations as compared with the bulk of the InN film (by one or even two orders of magnitude). The typical dislocation density near the InN/GaN interface is about $10^{11}$ cm$^{-2}$ that corresponds to the distance between dislocations of approximately 30 nm. This value is less or comparable with $\lambda$, which prevents the electrons in the near-interface layer to be quantized in the magnetic fields employed due to their fast scattering. Therefore these electrons cannot contribute to the SdH oscillations, providing the difference between the electron concentration measured in the bulk of the InN matrix (Hall concentration) and that determined from the SdH oscillations. This difference may serve as an estimate of the structural quality of the InN layer and/or the efficiency of the initial growth stage to suppress propagation of threading dislocations from the InN/GaN interface. It is worth noting that the similar difference was observed earlier \cite{14}, but was not explained. The only sample where the near-interface layer electrons seem to contribute to the SdH oscillations is sample E980 demonstrating three sets of oscillations (Fig. 1). The concentration of the electrons, defined from the period of intermediate SdH oscillations as $8.0\times10^{18}$cm$^{-3}$, and their mobility of $\sim$920 cm$^2$/Vs corresponding to the magnetic field of the onset of the oscillations fit well to transport parameters of the InN matrix presented in Table 1, in assumption of the conventional thickness of the near-interface layer of 100-200 nm and taking into account the bulk and surface layer contributions. The 3D near-interface electrons in this sample may participate in SdH oscillations in the $\bf{B}$$\parallel$$\bf{c}$-axis configuration probably due to the transformation of the near-interface extended defect structure, induced by high Mg doping, as reported by Liliental-Weber et al. for similar samples \cite{17}. According to \cite{17}, the significant Mg doping in the sample results in emergence of planar defects (stacking faults) separated from each other by $\sim$10 nm, which additionally reduce the density of other threading extended defects. On the other hand, the near-interface electrons are scattered efficiently by the planar defects, which results in damping of intermediate SdH oscillations in the $\bf{B}$$\perp$$\bf{c}$-axis configuration (Fig. 2).

In case of taking into account the total electron concentration in the bulk, comprising both quantized and near-interface electrons, the effect of the surface accumulation layer on the electrical measurements of the bulk InN parameters is expected to be even less pronounced.

\section{IV. Summary}
Four main contributions (In nanoparticles, surface, near-interface and bulk layers) to the conductivity of InN films grown by PA MBE have been found and identified. The ranges of the electron concentration and mobility in the surface accumulation layer have been determined directly for different undoped and Mg-doped InN films for the first time and shown to be $(1-3)\times10^{13}$cm$^{-2}$ and (400-600)cm$^2$/Vs, respectively. It has been established that the surface layer has no significant influence on the electrical measurements of high-mobility InN films, while for low-mobility compensated films (e.g. strongly Mg-doped) its effect can be pronounced. The observed difference between the Hall electron concentration and that of quantized bulk electrons has been explained by the influence of the near-interface layer usually containing much higher threading dislocation density.


%



\begin{acknowledgments}
The work has been supported in part by Programs of the Presidium of RAS, RFBR grant $\sharp$10-02-00689-a, and EuroMagNET II program under the EU contract $\sharp$228043.
\end{acknowledgments}


\end{document}